\documentclass[preprint,showpacs,preprintnumbers,aps,amsmath,amssymb]{revtex4-1}
\usepackage{graphicx,dcolumn,bm}
\usepackage[bf,SL,BF]{subfigure}
\usepackage{txfonts}

\begin{document}

\title{Ballistic Thermal Rectification in Asymmetric Three-Terminal Mesoscopic Dielectric Systems}

\author{Yi Ming }
\email{meanyee@mail.ustc.edu.cn}
 \affiliation{School of Physics and Material Science, Anhui
 University,\\
 Hefei, Anhui 230039, People's Republic of China}

\author{Zhe Xian Wang}
\author{Ze Jun Ding}
\affiliation{
Hefei National Laboratory for Physical Sciences at Microscale }
\affiliation{
Department of Physics, University of Science and Technology of
China,\\
Hefei, Anhui 230026, People's Republic of China
}%
\author{Hui Min Li} 
\affiliation{
Qingdao Institute of Bioenergy and Bioprocess Technology, Chinese Academy of Sciences,\\
Qingdao, Shandong 266101, People's Republic of China
}

\date{\today}
\begin{abstract}
By coupling the asymmetric three-terminal mesoscopic dielectric system with a temperature probe, at low temperature, the ballistic heat flux flow through the other two asymmetric terminals in the nonlinear response regime is studied based on the Landauer formulation of transport theory. The thermal rectification is attained at the quantum regime. It is a purely quantum effect and is determined by the dependence of the ratio $\tau_{RC}(\omega)/\tau_{RL}(\omega)$ on $\omega$, the  phonon's frequency. Where $\tau_{RC}(\omega)$ and $\tau_{RL}(\omega)$ are respectively the transmission coefficients from two asymmetric terminals to the temperature probe, which are determined by the inelastic scattering of ballistic phonons in the temperature probe. Our results are confirmed by extensive numerical simulations.
\end{abstract}
\maketitle

At low temperature, the thermal conductance in mesoscopic phonon systems is dominated by the transmission of ballistic phonons.
Using the Landauer formulation of transport theory, Rego and Kirczenow\cite{Rego98} predicted the existence of the
quantum of ballistic thermal conductance. Further investigations indicated that the quantum of ballistic thermal conductance is universal and it is independent of the statistics obeyed by the carriers\cite{Rego99,Blencowe00,Krive99}. The quantum of ballistic thermal conductance has already been verified by experiment\cite{Schwab00} and the universality is verified later\cite{Meschke06}. Since then, the quantum of ballistic thermal conductance has attracted much attention\cite{Sun02,Mingo05,Chiu05,Yamamoto06,Chiatti06}. Besides the unique property of the quantized thermal conductance in the mososcopic systems, the other motivation for these investigations is to seek ways to control the heat transport at low temperature.

To control the heat transport, the first nanoscale model for thermal rectifier was proposed in 1D nonlinear chains in 2002\cite{Terraneo02}. After that, many theoretical works were performed to study the thermal rectification\cite{Li04,Li05,Li06,Hu06,Segal05,Segal08,Wu09,Eckmann06,Casati07}, and the first experimental
work to demonstrate the thermal rectification was also reported\cite{Chang06}, although only the $2$ to $7\%$ rectification efficiency is yielded. Recently, based on the 1D nonlinear chains, the models of thermal logic gates\cite{Wang07} and thermal memory\cite{Wang08} were proposed to demonstrate that phonons can be used to carry information and processed accordingly. Although considerable achievements in thermal rectification based on the nonlinear crystal are made, the thermal rectification based on the harmonic system is still unavailable. This may due to the fact that there is no thermal rectification in the heat transport in a two-terminal harmonic system at low temperature\cite{Segal05}. This absence of thermal rectification in harmonic system is also found in classical models: the harmonic chains with self-consistent reservoirs\cite{Pereira08,Segal09} and harmonic $N$-terminal junction\cite{Segal08b}. However, at the quantum regime, for an asymmetric harmonic system which couples with a temperature probe connecting directly to a self-consistent reservoir, the temperatures of the self-consistent reservoir for forward and reversed
bias are not symmetric with respect to the average temperature\cite{Ming07}. Therefore, one can expect that the thermal rectification can be attained in this asymmetric harmonic system at the quantum regime.

In this work, we theoretically demonstrate the possibility to attain the ballistic thermal rectification by coupling the asymmetric mesoscopic three-terminal junctions with a temperature probe. This possibility is confirmed by extensive numerical simulations. Just same as the voltage probe in the mesoscopic electronic systems\cite{Buttiker86,Buttiker88}, which has been used to study the electrical rectification in three-terminal electrical ballistic junctions\cite{Jordan08}, the temperature probe acts as a dephasing probe which introduces inelastic scattering of phonons in three-terminal junction. This inelastic scattering brings nonlinearity in the asymmetric three-terminal junctions and makes the thermal rectification possible.

\begin{center}
\begin{figure}[htbp]
\includegraphics[width=0.5\textwidth]{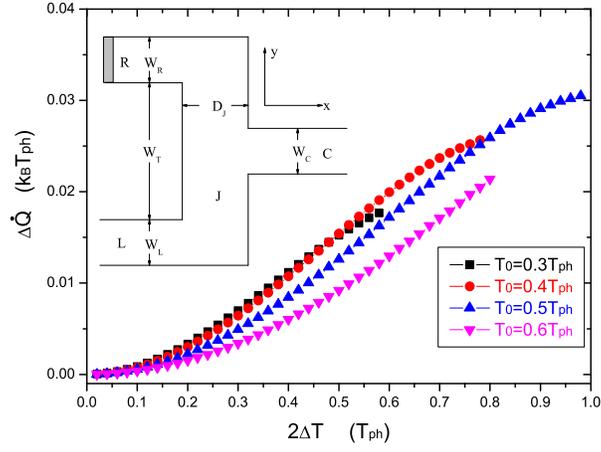}
\caption{\label{fig1}(color online) The difference of heat flux $\Delta\dot{Q}=\dot{Q}_{-}+\dot{Q}_{+}$ (in the unit of $k_BT_{ph}$) vs $2\Delta T$ ($=T_{hot}-T_{cold}$) at different $T_0$, where $W_T=21$~nm. Inset: schematic illustration of an asymmetric
three-terminal mesoscopic dielectric system coupled a temperature probe by the $R$ terminal.}
\end{figure}
\end{center}

The asymmetric three-terminal junction is sketched in the inset of Fig.~\ref{fig1}. Through terminals $L$ and $C$, the junction is coupled to two thermal reservoirs at temperatures $T_L$ and $T_C$. The steady-state heat flux $\dot{Q}\equiv\dot{Q_C}=-\dot{Q_L}$ is passed through the junction via terminal $C$ and $L$. A third terminal $R$ is connected to another thermal reservoir at temperature $T_R$. This third terminal is a temperature probe, it means $T_R$ is adjusted in such a way that no net heat flux pass through terminal $R$. The ballistic regime is considered in this work. The heat flux $\dot{Q}_i$ from terminal $i$ ($i=L,R,C$)
flowing into the midsection $J$ can be expressed as
\cite{Sun02,Yang04}
\begin{equation}\label{eqn1}
    \dot{Q}_i=\sum_{j (j\neq i)} \int_0^{+\infty}
    [n(T_i,\omega)-n(T_j,\omega)]\hbar\omega\, \tau_{ji}(\omega) \frac{d\omega}{2\pi},
\end{equation}
where $n(T_i,\omega)=[\exp (\hbar \omega/k_B T_i)-1]^{-1}$ is the
Bose-Einstein distribution function of the phonons in the $i$th
reservoir, $T_i$ is the equilibrium temperature of thermal reservoir
$i$, $\tau_{ji}(\omega)=\sum_m \theta (\omega -\omega_{im})\tau_{ji,m}(\omega)=\sum_{m,n} \theta (\omega -\omega_{jn}) \theta (\omega -\omega_{im})
\tau_{ji,nm}(\omega)$ is the total transmission coefficient and $\tau_{ji,nm}(\omega)$ is the transmission coefficient
from mode $m$ of terminal $i$ at frequency $\omega$ across all the
interface into the mode $n$ of terminal $j$, $\omega_{im}$ is the cutoff frequency of mode $m$ in terminal
$i$. By the time-reversal symmetry, we have $\tau_{ji}(\omega)=\tau_{ij}(\omega)$. The influence of the temperature probe $R$ on the steady-state heat flux $\dot{Q}$ is determined by the transmission coefficients $\tau_{RC}(\omega)$ and $\tau_{RL}(\omega)$.

Firstly, the reservoir $L$ is set at temperature $T_L=T_{hot}$, which is higher than the temperature $T_C=T_{cold}$ of reservoir $C$. The heat flux flows through the system is $\dot{Q}_{-}\equiv\dot{Q_C}$.
Evidently, the heat flux flows into $C$ reservoir hence $\dot{Q}_-<0$. By reversing the temperature bias, i.e., let $T_L=T_{cold}$ and $T_C=T_{hot}$, the heat flux equals to $\dot{Q}_{+}\equiv\dot{Q}_C>0$.

To study whether the ballistic thermal rectification will exhibit, $\dot{Q}_{-}$ is added to $\dot{Q}_{+}$.
\begin{eqnarray}\label{eqn4}
&&\Delta \dot{Q}\equiv \dot{Q}_{-}+\dot{Q}_{+}=
\int_0^{+\infty}[n(T_{cold},\omega)-n(T_R,\omega)\nonumber\\
&&+n(T_{hot},\omega)-n(T'_R,\omega)]\tau_{RC}(\omega)\hbar\omega\,  \frac{d\omega}{2\pi},
\end{eqnarray}
where $T_R$ and $T'_R$ are determined by $\dot{Q}_R=0$. Obviously, $T_R$ and $T'_R$ are determined by the transmission coefficients $\tau_{RC}(\omega)$ and $\tau_{RL}(\omega)$. Therefore, whether the rectification can be attained depends critically on the transmission coefficients $\tau_{RC}(\omega)$ and $\tau_{RL}(\omega)$.
In the linear response regime, by using $\dot{Q}_i=\sum_{j\neq i}G_{ji}(T_i-T_j)$\cite{Sun02,Yang04}, where $G_{ji}$ is the two-terminal thermal conductance from reservoir $i$ to $j$, one can easily find that $n(T_{cold},\omega)-n(T_R,\omega)+n(T_{hot},\omega)-n(T'_R,\omega)=0$ for any $\omega$. Thus $\Delta \dot{Q}=0$, which means there isn't rectification in the linear response regime. This corresponds to the one obtained in \cite{Jacquet09} for the thermoelectric transport in a chain of quantum dots with self-consistent reservoirs. On the other hand, if the ratio $\tau_{RC}(\omega)/\tau_{RL}(\omega)$ is a constant for any frequency $\omega$, by using $\dot{Q}_R=0$, one can easily find $\Delta\dot{Q}=0$ and the rectification is absent even in the nonlinear response regime. Therefore, even in the nonlinear regime, the thermal rectification is absent in the symmetric three-terminal junctions with $\tau_{RC}(\omega)=\tau_{RL}(\omega)$.

In a more general situation where the ratio $\tau_{RC}(\omega)/\tau_{RL}(\omega)$ for any frequency is not a constant, the fact that $\tau_{RC}(\omega)/\tau_{RL}(\omega)$ varies with $\omega$ means there are different asymmetries for transmission of phonons in different frequencies. One can certainly expect that $\Delta\dot{Q}$ is not zero in the nonlinear response regime and it is determined by the dependence of the ratio $\tau_{RC}(\omega)/\tau_{RL}(\omega)$ on $\omega$. We start with letting $T_0=(T_{hot}+T_{cold})/2$ and $\Delta T=(T_{hot}-T_{cold})/2$, where $T_0$ can be regarded as the working temperature according to the following text. When $\Delta T$ is a small value, by using the Taylor
expansion of the Bose-Einstein distribution function $n(T_i,\omega)$ as well as the approximate results of $T_R$ and $T'_R$ in ref.~\cite{Ming07} for the asymmetric three-terminal junctions, one can find\footnote{We appreciate Dr. Philippe A. Jacquet for his helpful correction on the factor $(1-\beta^2)$.}
\begin{eqnarray}\label{eqn5}
\Delta\dot{Q}&=&(1-\beta^2)\frac{F_2(\tau_{RC})F_1(\tau_{RL})-F_2(\tau_{RL})F_1(\tau_{RC})}{F_1(\tau_{RC})+F_1(\tau_{RL})}(\Delta T)^2\nonumber\\
&&+\mathcal {O}
[(\Delta T)^4]\equiv\alpha(\Delta T)^2+\mathcal {O}
[(\Delta T)^4].
\end{eqnarray}
where
\begin{eqnarray}
F_k(\tau_{Ri})&=&\int_0^{+\infty}\left( \frac{\partial^k n}
{\partial T^k} \right)_{T_0}\hbar \omega \tau_{Ri}
\frac{d\omega}{2\pi},\label{eqn5a}\\
\beta&=&\frac{F_1(\tau_{RC})-F_1(\tau_{RL})}{F_1(\tau_{RC})+F_1(\tau_{RL})}.\label{eqn5b}
\end{eqnarray}
By a simple algebra, $F_2(\tau_{RC})F_1(\tau_{RL})-F_2(\tau_{RL})F_1(\tau_{RC})$ can be reexpressed as
\begin{eqnarray}\label{main}
&&F_2(\tau_{RC})F_1(\tau_{RL})-F_2(\tau_{RL})F_1(\tau_{RC})\nonumber\\
&=&\int_0^{+\infty}\left[ \frac{\partial^2 n(\omega_1)}
{\partial T^2} \right]_{T_0}\hbar \omega_1 \tau_{RC}(\omega_1)
\frac{d\omega_1}{2\pi}\int_0^{+\infty}\left[ \frac{\partial n(\omega_2)}
{\partial T} \right]_{T_0}\hbar \omega_2 \tau_{RL}(\omega_2)
\frac{d\omega_2}{2\pi}\nonumber\\
&&-\int_0^{+\infty}\left[ \frac{\partial^2 n(\omega_1)}
{\partial T^2} \right]_{T_0}\hbar \omega_1 \tau_{RL}(\omega_1)
\frac{d\omega_1}{2\pi}\int_0^{+\infty}\left[ \frac{\partial n(\omega_2)}
{\partial T} \right]_{T_0}\hbar \omega_2 \tau_{RC}(\omega_2)
\frac{d\omega_2}{2\pi}\nonumber\\
&=&\int_0^{+\infty} d\omega_2 \int_0^{+\infty} d\omega_1 \left[ \frac{\partial^2 n(\omega_1)}
{\partial T^2} \right]_{T_0} \left[ \frac{\partial n(\omega_2)}
{\partial T} \right]_{T_0}\mathcal{T}(\omega_1,\omega_2)\hbar\omega_1 \hbar\omega_2/(2\pi)^2\nonumber\\
&=&\int_0^{+\infty} d\omega_2 \int_0^{\omega_2} d\omega_1 \left[ \frac{\partial^2 n(\omega_1)}
{\partial T^2} \right]_{T_0} \left[ \frac{\partial n(\omega_2)}
{\partial T} \right]_{T_0}\mathcal{T}(\omega_1,\omega_2)\hbar\omega_1 \hbar\omega_2/(2\pi)^2\nonumber\\
&&+\int_0^{+\infty} d\omega_2 \int_{\omega_2}^{+\infty} d\omega_1 \left[ \frac{\partial^2 n(\omega_1)}
{\partial T^2} \right]_{T_0} \left[ \frac{\partial n(\omega_2)}
{\partial T} \right]_{T_0}\mathcal{T}(\omega_1,\omega_2)\hbar\omega_1 \hbar\omega_2/(2\pi)^2\nonumber\\
&=&\int_0^{+\infty} d\omega_1 \int_0^{\omega_1} d\omega_2 \left[ \frac{\partial^2 n(\omega_2)}
{\partial T^2} \right]_{T_0} \left[ \frac{\partial n(\omega_1)}
{\partial T} \right]_{T_0}\mathcal{T}(\omega_2,\omega_1)\hbar\omega_2 \hbar\omega_1/(2\pi)^2\nonumber\\
&&+\int_0^{+\infty} d\omega_2 \int_{\omega_2}^{+\infty} d\omega_1 \left[ \frac{\partial^2 n(\omega_1)}
{\partial T^2} \right]_{T_0} \left[ \frac{\partial n(\omega_2)}
{\partial T} \right]_{T_0}\mathcal{T}(\omega_1,\omega_2)\hbar\omega_1 \hbar\omega_2/(2\pi)^2\nonumber\\
&=&\int_0^{+\infty} d\omega_2 \int_{\omega_2}^{+\infty}d\omega_1\mathcal {N}(\omega_1,\omega_2,T_0)\mathcal {T}(\omega_1,\omega_2)\hbar\omega_1 \hbar\omega_2/(2\pi)^2,
\end{eqnarray}
where
\begin{eqnarray}
\mathcal {N}(\omega_1,\omega_2,T_0)&=&\left[\frac{\partial^2n(\omega_1)}{\partial T^2}\frac{\partial n(\omega_2)}{\partial T} -\frac{\partial^2n(\omega_2)}{\partial T^2}\frac{\partial n(\omega_1)}{\partial T}\right]_{T_0}\nonumber\\
&=&\left\{\frac{\hbar\omega_1}{k_BT_0}[ 1+2n(\omega_1)]-\frac{\hbar\omega_2}{k_BT_0}[ 1+2n(\omega_2)]\right\}_{T_0}\nonumber\\
&&\times\frac{1}{T_0}\left[\frac{\partial n(\omega_1)}{\partial T}\frac{\partial n(\omega_2)}{\partial T}\right]_{T_0},\label{eqn5c}\\
\mathcal {T}(\omega_1,\omega_2)&=&-\mathcal {T}(\omega_2,\omega_1)=\tau_{RC}(\omega_1)\tau_{RL}(\omega_2)-\tau_{RC}(\omega_2)\tau_{RL}(\omega_1).\label{eqn5d}
\end{eqnarray}
This is the main result of this work.
In Eq.~(\ref{eqn5c}), $\partial^2 n/\partial T^2=[(1+2n)\hbar\omega/(k_B T^2)-2/T]\partial n/\partial T$ is used. Obviously, $\mathcal{N} (\omega_1,\omega_2,T_0)$ is $T_0$ dependent and is originating from the quantum statistics of phonons. $\mathcal {T}(\omega_1,\omega_2)$ is originating from the inelastic scattering of phonons in the temperature probe. The first term of $\mathcal{T}(\omega_1,\omega_2)$ in Eq.~(\ref{eqn5d}) can be understood as that the ballistic phonon with frequency $\omega_1$ transport from reservoir $C$ into temperature probe $R$ with transmission coefficient $\tau_{RC}(\omega_1)$, after being inelastically scattered into the different frequency $\omega_2$, the ballistic phonon transport from the temperature probe into reservoir $L$ with transmission coefficient $\tau_{RL}(\omega_2)$. The second term presents the similar process but in the inverse direction. Thus the ballistic thermal rectification is a purely quantum effect and is determined by the inelastic scattering in the temperature probe. To study the sign of $\mathcal{N}(\omega_1,\omega_2,T_0)$, $\partial\{x[1+2n(x)]\}/\partial x$ is studied, where $x\equiv\hbar\omega/k_BT_0$.
\begin{eqnarray}
\frac{\partial\{x[1+2n(x)]\}}{\partial x}&=&\frac{e^{2x}-2xe^x-1}{(e^x-1)^2}\nonumber\\
&=&\frac{1}{(e^x-1)^2}\sum_{l=2}^\infty\frac{2^l-l-1}{(l+1)!}2x^{l+1}>0\label{eqn5e}
\end{eqnarray}
for all $x>0$.
Hence $\mathcal{N}(\omega_1,\omega_2,T_0)>0$ for all $\omega_1>\omega_2>0$. Thus, the inelastic scattering factor, $\mathcal {T}(\omega_1,\omega_2)$, determines the sign of $\alpha$ due to $0\leq\beta^2<1$. By reexpressed $\mathcal{T}(\omega_1,\omega_2)$ as $[\tau_{RC}(\omega_1)/ \tau_{RL}(\omega_1)-\tau_{RC}(\omega_2)/ \tau_{RL}(\omega_2)]\tau_{RL}(\omega_1)\tau_{RL}(\omega_2)$ when $\tau_{RL}(\omega)\neq 0$ for all $\omega$, one can easily find that the sign of $\alpha$ is determined by the dependence of the ratio $\tau_{RC}(\omega)/\tau_{RL}(\omega)$ on $\omega$. $\Delta\dot{Q}>0$ when $\tau_{RC}(\omega)/\tau_{RL}(\omega)$ increases with $\omega$.

To obtain the exact $\Delta\dot{Q}$, we carry out numerical calculations for Eq.~(\ref{eqn4}) with the scattering matrix method used in \cite{Ming07}. In the
calculation, the geometrical parameters of the system are chosen as $W_L=W_R=W_C=D_J=10$~nm while $W_T$ can be varied. We
limit the temperature $T_{hot}$ of the thermal reservoir at the higher temperature is lower than $T_{ph} =\hbar \pi v/W_L k_B \approx 7.61$~K
($v$ is the sound velocity) to ensure that the phonon relaxation can be neglected \cite{Sun02} and
the heat conduction is determined by the ballistic
transmission of the acoustic phonons.

\begin{center}
\begin{figure}[htbp]
\includegraphics[width=0.5\textwidth]{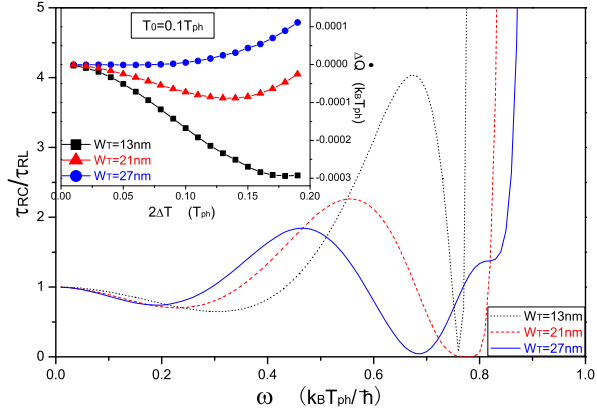}
\caption{\label{fig2}(color online) The dependence of the ratio $\tau_{RC}/\tau_{RL}$ on the frequency $\omega$ (in the unit of $k_BT_{ph}/\hbar$). Inset: $\Delta\dot{Q}$ vs $2\Delta T$ for different $W_T$'s at the average temperatures of $T_0=0.1T_{ph}$.}
\end{figure}
\end{center}

Fig.~\ref{fig1} shows the results of $\Delta\dot{Q}$ vs $2\Delta T$ for different $T_0$ with $W_T=21$~nm. Firstly, at small but finite $\Delta T$, $\Delta\dot{Q}$ shows a quadratic dependence
on $\Delta T$, in agreement with Eq.~(\ref{eqn5}). Secondly, corresponds to the factor $\mathcal{N}(\omega_1,\omega_2,T_0)$, $\Delta\dot{Q}$ decreases with increasing $T_0$ for a system with a certain $W_T$. Thirdly, when the temperature difference $2\Delta T=T_{hot}-T_{cold}$ is finite, the thermal rectification is obtained with $\Delta\dot{Q}> 0$ in Fig.~\ref{fig1}. It means $\mathcal{T}(\omega_1,\omega_2)>0$ at these $T_0$'s hence there is a greater heat flux in the direction from $C$ reservoir to $L$ reservoir than the heat flux in the inverse direction when the temperature bias is reversed.
This indicates that by suffering the inelastic scattering introduced by the temperature probe, ballistic phonon has more probability to transport from $C$ reservoir into $L$ reservoir than transport in the inverse direction. To understand this, one must recall the fact that $\Delta\dot{Q}$ is determined by the factors $\mathcal{N}(\omega_1,\omega_2,T_0)$ and $\mathcal{T}(\omega_1,\omega_2)$ as shown in Eq.~(\ref{main}). The phonons in reservoirs obey the Bose-Einstein statistics, as well as the fact that $\partial\{x[1+2n(x)]\}/\partial x\to 0$ when $x\to 0$ as shown in Eq.~(\ref{eqn5e}), the phonons with the frequencies in the region $0<\omega<3T_0/T_{ph}$ (where the unit, $k_BT_{ph}/\hbar$, is omitted) dominate $\mathcal{N}(\omega_1,\omega_2,T_0)$ when $\Delta T$ is a small finite value and the phonon with the lower frequency is more dominant. Thus for $W_T=21$~nm, at $T_0=0.3T_{ph}$, the phonons with frequencies $0<\omega< 0.9$ dominate $\mathcal{N}(\omega_1,\omega_2,T_0)$. The phonons with frequencies $0<\omega< 0.25$ is more dominant to $\mathcal{N}(\omega_1,\omega_2,T_0)$, but as shown in Fig.~\ref{fig2}, $\tau_{RC}/\tau_{RL}$ decreases much slowly in this region and it increases much faster in the region $0.25<\omega< 0.58$. Hence by considering the both contributions from $\mathcal{N}(\omega_1,\omega_2,T_0)$ and $\mathcal{T}(\omega_1,\omega_2)$, the phonons with frequencies $0.25<\omega< 0.58$ determine $\Delta\dot{Q}>0$ because $\mathcal{T}(\omega_1,\omega_2)>0$ in this region. The similar analysis can be applied to other $T_0$'s. One can expect that at very low $T_0$, $\Delta\dot{Q}<0$ because $\mathcal{T}(\omega_1,\omega_2)<0$ in the dominant region of frequency. It is confirmed by the inset of Fig.~\ref{fig2} at $T_0=0.1T_{ph}$.

\begin{center}
\begin{figure}[htbp]
\includegraphics[width=0.5\textwidth]{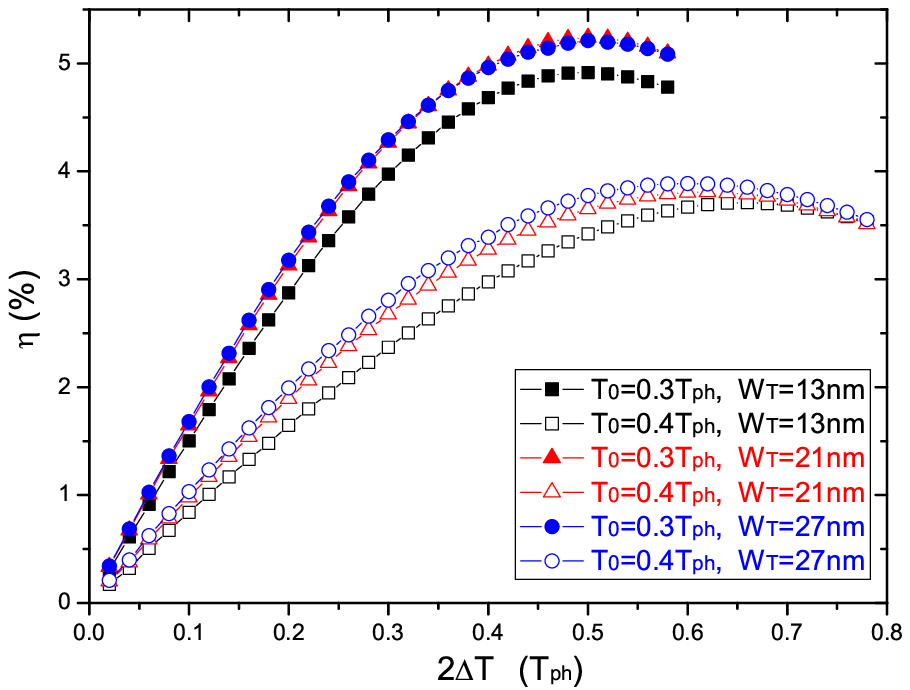}
\caption{\label{fig3}(color online) The efficiency of rectification vs the temperature difference $2\Delta T$ ($=T_{hot}-T_{cold}$) for different $W_T$'s at the average temperatures of $T_0=0.3T_{ph}$ and $0.4T_{ph}$.}
\end{figure}
\end{center}

\begin{center}
\begin{figure}[htbp]
\includegraphics[width=0.5\textwidth]{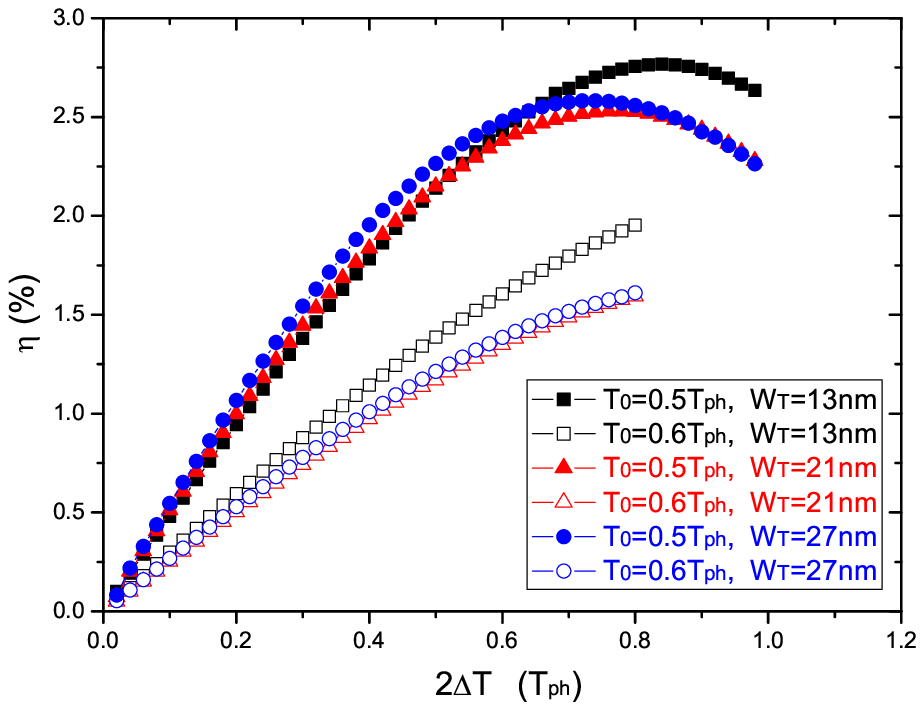}
\caption{\label{fig4}(color online) The efficiency of rectification vs the temperature difference $2\Delta T$ ($=T_{hot}-T_{cold}$) for different $W_T$'s at the average temperatures of $T_0=0.5T_{ph}$ and $0.6T_{ph}$.}
\end{figure}
\end{center}

The thermal rectification can be studied by the efficiency which is defined as $\eta=|\Delta\dot{Q}|/{|\dot{Q}_-|}\times 100\%$.
The numerical results of the efficiency are shown in Figs.~\ref{fig3} and~\ref{fig4} for different $T_0$'s and different  $W_T$'s. The efficiency can achieve about $5.5\%$ as shown in the two figures. For the system with $W_T=27$~nm, the region of frequency in which $\tau_{RC}/\tau_{RL}$ slowly decreases is the shortest as shown in Fig.~\ref{fig2}. Thus at $T_0=0.3T_{ph}$, $0.4T_{ph}$ and $0.5T_{ph}$, this system achieves the highest efficiency. However, at $T_0=0.6T_{ph}$, the system with $W_T=13$~nm has the highest efficiency. This is because $\tau_{RC}/\tau_{RL}$ increases in the region $0.35<\omega< 0.7$.

In summary, we have studied the ballistic phonon heat flux flows in the asymmetric three-terminal systems and the ballistic thermal rectification is attained. The rectification is purely a quantum effect and is critically determined by the dependence of the ratio $\tau_{RC}(\omega)/\tau_{RL}(\omega)$ on $\omega$. $\tau_{RC}(\omega)/\tau_{RL}(\omega)$ is determined by the inelastic scattering of phonons in the temperature probe. Therefore, one can attain the ballistic thermal rectification by introducing the inelastic scattering of phonons into the mesoscopic systems.

\begin{acknowledgments}
Yi Ming is supported by the Anhui Provincial Natural Science Foundation (Grant No.~090416235). Ze Jun Ding is supported by the National Natural Science Foundation of China (Grant No.~10874160), `111' project, Chinese Education Ministry and Chinese Academy of Sciences.
\end{acknowledgments}

%

\end{document}